\documentclass[12pt]{iopart}

\renewcommand{\fnsymbol}[1]{^\arabic{footnote}}

\newcommand{\beq}{\begin{equation}}
\newcommand{\eeq}{\end{equation}}
\newcommand{\beqn}{\begin{eqnarray}}
\newcommand{\eeqn}{\end{eqnarray}}

\newcommand{\sunmass}{M_{\odot}}

\def\eadnew#1#2{\address{#2 E-mail: \mailto{#1}}}

\expandafter\let\csname equation*\endcsname\relax
\expandafter\let\csname endequation*\endcsname\relax
\usepackage{amsmath}
\usepackage{amssymb}

\usepackage{enumerate}
\usepackage{ellipsis}
\usepackage{afterpage}
\usepackage{hhline}
\usepackage{multirow}
\usepackage{graphicx}
\usepackage{epsf}
\usepackage{longtable}
\usepackage{hyperref}
\usepackage{graphics,epsfig,placeins,wrapfig}
\usepackage{subcaption}
\usepackage[usenames]{color}
\usepackage{soul}
\usepackage{floatrow}
\usepackage{cite}

\floatstyle{plaintop}
\restylefloat{table}

\newenvironment{packed_enumerate}{
\begin{enumerate}[1.]
  \setlength{\itemsep}{1pt}
  \setlength{\parskip}{0pt}
  \setlength{\parsep}{0pt}
}{\end{enumerate}}

\begin{document}
\title{Optimizing spinning time-domain gravitational waveforms for Advanced LIGO data analysis \hspace{0.5cm}}

\author{
  Caleb Devine$^{1,*}$,
  Zachariah B.~Etienne$^{1,\dag}$
  Sean T.~McWilliams$^{2,\ddag}$
  }
\address{$^1$ Department of Mathematics, West Virginia University, Morgantown, WV 26506, USA}
\address{$^2$ Department of Physics and Astronomy, West Virginia University, Morgantown, WV 26506, USA}
\eadnew{rdevine1@mix.wvu.edu}{$^{*}$}
\eadnew{zbetienne@mail.wvu.edu}{$^{\dag}$}
\eadnew{Sean.McWilliams@mail.wvu.edu}{$^{\ddag}$}

\begin{abstract}
The Spinning Effective One Body-Numerical Relativity (SEOBNR) series
of gravitational wave approximants are among the best available for
Advanced LIGO data analysis. Unfortunately, SEOBNR codes as they
currently exist within LALSuite are generally too slow to be directly
useful for standard Markov-Chain Monte Carlo-based parameter
estimation (PE). Reduced-Order Models (ROMs) of SEOBNR have been
developed for this purpose, but there is no known way to make ROMs of
the full eight-dimensional intrinsic parameter space more efficient
for PE than the SEOBNR codes directly. So as a proof of principle, we
have sped up the original LALSuite SEOBNRv2 approximant code, which
models waveforms from aligned-spin systems, by nearly 300x. Our
optimized code shortens the timescale for conducting PE with this
approximant to months, assuming a purely serial analysis, so that even
modest parallelization combined with our optimized code will make
running the full PE pipeline with SEOBNR codes directly a realistic
possibility. A number of our SEOBNRv2 optimizations have already been
applied to SEOBNRv3, a new approximant capable of modeling sources
with all eight (precessing) intrinsic degrees of freedom. We
anticipate that once all of our optimizations have been applied to
SEOBNRv3, a similar speed-up may be achieved. 

 \end{abstract}


\maketitle

\section{Introduction}
The Advanced Laser Interferometer Gravitational-wave
Observatory~\cite{AdvLIGO} (LIGO) and the French-Italian gravitational
wave detector, VIRGO,~\cite{virgo} are the most sensitive
gravitational-wave interferometers that have ever been constructed.
Although Initial/Enhanced LIGO/VIRGO did not detect gravitational
waves~\cite{nowavesenhanced}, Advanced LIGO~\cite{AdvLIGO} achieved unprecedented
sensitivities during its first observing run 
and detected gravitational waves for the first time~\cite{DetectionPaper2016}.

Spins and masses of compact objects can be inferred from detected
gravitational waves through a Bayesian model approximation using
Markov chain Monte Carlo (MCMC) methods implemented in
{\sc  lalinference}~\cite{Veitch2015}. 
Ideally, the theoretical waveforms required for these methods will be
based entirely on {\it full} solutions to Einstein's equations
(numerical relativity). On the surface this seems promising, as
numerical relativity work can reliably solve the full set of
Einstein's equations for all compact binary systems of interest for
Advanced LIGO, and can even generate gravitational waves for long
inspirals. For example, state-of-the-art numerical relativity
simulations have successfully generated a 350-cycle black-hole binary
gravitational waveform that spans {\it the entire LIGO band} for
systems with total mass over
$45.5\sunmass$~\cite{Bela2015}. Unfortunately, generation of this
single waveform spanned {\it months} on a high-performance computing
resource, and traditional MCMC parameter estimation (PE) across the
full space of possible binary parameters will require $\sim10^8$
waveforms to be generated {\it sequentially}. Therefore, full
parameter estimation for a single observed wave using current
numerical relativity techniques and resources could require $\sim$5
million years, or roughly 7 orders of magnitude too long to be useful
for LSC work.

There are a number of strategies for overcoming this enormous
computational challenge, generally relying on perturbative and/or
phenomenological solutions to Einstein's equations. Since numerical
relativity directly solves the full set of Einstein’s equations, it
can be argued that the most accurate approximants will incorporate
numerical relativity solutions. However, all approximants possess
their own set of systematic errors.

As an example, consider the ``Phenom'' family of
phenomenological gravitational waveform
models~\cite{Hannam2014,Husa2015a,Husa2015b,Kumar2016}. Recent ``Phenom'' models
use the state-of-the-art, purely perturbative SEOBv2 
(Spinning Effective One-Body, version 2) model
for the inspiral part of the waveform (i.e., an uncalibrated version
of the~\cite{Taracchini2014} model), 
and for merger and ringdown,
attach a phenomenological waveform calibrated to $\sim 20$ numerical
relativity waveforms~\cite{Husa2015b}.

``Phenom'' models have the great advantage of being able to
generate theoretical waveforms extremely fast and in the frequency
domain directly, which simplifies most data analyses. As a result,
they are one of the bedrocks of LSC parameter estimation.
That being said, both ``Phenom'' and SEOBNR models possess unique
systematic uncertainties that are magnified by the fact that few
numerical relativity waveforms exist that are sufficiently long to
fully uncover systematic errors in approximate models across the full
parameter space. {\it Therefore, there is a strong need for fully
  independent gravitational wave approximants with different
  systematic uncertainties that are capable of spanning the widest
  parameter space possible.}
Perhaps more importantly, given the
significantly different approaches used to model the signal using each
technique, there is good reason to hope that their systematic
uncertainties will be independent of one another, so they can be
used to better understand the overall systematic uncertainties
resulting from waveform modeling errors.

The SEOBNR series of waveform
models~\cite{Buonanno2007,Damour2008,Damour2009,Pan2010,Yunes2010,Pan2011,Taracchini2012,Damour2013,Pan2014,Taracchini2014,Nagar2015}
fill this gap extremely well.
Although the calculation of the inspiral by SEOBNRv2 and the most 
recent version of Phenom are not entirely independent~\cite{Husa2015b},
the merger and ringdown are entirely independent.
The SEOBNR approximant,~\cite{Taracchini2014}, which is currently on its
second version (SEOBNRv2), contains much of the relevant physics,
including spin-orbit effects up to order 3.5PN~\cite{eobnr35PN}, and
the ability to model varying mass ratios and individual spin
magnitudes for each compact object; it lacks only the ability to model
spins that are not aligned with the orbital angular momentum, and
therefore precess over time. This feature has only now been added to
the third version (SEOBNRv3). 

Like ``Phenom'' models, SEOBNR models have been shown to produce waveforms
that agree well with some of the latest numerical relativity 
black-hole binary waveform calculations. Though SEOBNR is truly a
state-of-the-art approximant, like any approximant it possesses systematic
uncertainties, again magnified by the fact that there are so few numerical
relativity waveforms against which it could be calibrated, particularly
for systems with precessing spins.
Despite their very strong pedigree, SEOBNR waveform models---as they
have been officially implemented in {\tt LALSuite}~\cite{LALSuite:web}
(the LSC's open-source data analysis software repository)---are
extremely slow, requiring roughly six minutes to generate a single binary
neutron star waveform that spans the entire Advanced LIGO band, given a 
start frequency of 10 Hz and a waveform sample rate set to 16,384 Hz,
which is the LALSuite default output frequency\footnote{This output
  frequency was also used in benchmarking the SEOBNRv2 ROM code in
  \cite{Purrer2014}.}.
SEOBNRv2-based parameter estimation using standard MCMC, which
requires the generation of $\sim 10^7$ sequential
waveforms~\cite{Brown2012}, would therefore take $\sim$100 {\it years}
to complete using current computer hardware. The time required
increases to a {\it millennium} for the $\sim 10^8$ sequential
waveforms necessary for parameter estimation across the full eight
dimensions of intrinsic black-hole binary parameter
space\footnote{Note that eight dimensions includes
  the mass ratio, $\mu$, each spin component~\cite{Hannam2014},
  and the total mass, where the total mass is only a scaling
  factor. Excluding the total mass, we are left with seven dimensions
  for SEOBNRv3 and three dimensions for SEOBNRv2. Regardless of how
  the dimensionality is tallied, the total mass must be considered for
  PE, so we quote four/eight dimensions for
  spin-aligned/spin-precessing binaries.}~\cite{Veitch2015}. As we will show, since 
such a large number of sequential waveforms generations are required,
our optimizations are absolutely essential to making possible a full
eight-dimensional pipeline using SEOBNR codes directly.

As evidence of the importance of the SEOBNR series, an ``industry''
has been built up within the gravitational-wave data-analysis
community around constructing Reduced Order Models (ROMs) of SEOBNR
waveforms (e.g.,~\cite{Field2014,Purrer2014,Purrer2015}). ROMs are 
built from a large sample of computed waveforms from SEOBNR models,
and take advantage of the fact that a small perturbation in initial
binary parameters will result in a small perturbation in the
resulting waveform; reduced bases and
interpolated parameter dependent coefficients are integral building
blocks in the construction of ROMs. In short, ROMs generate waveforms
between sampled points using what amounts to multidimensional
interpolations, though this may introduce small interpolation errors.

ROMs of SEOBNR models have been hugely successful so far, resulting in
speed-up factors of $\sim10^4$ over the original SEOBNR codes within
{\tt LALSuite}\footnote{To download source codes used to generate
results in this paper, first clone the latest development LALSuite git
repository, then run ``git checkout 7223f6e3a''}~\cite{Purrer2014,
Purrer2015}. Unfortunately, by their nature ROMs would need to be
regenerated if any recalibration to the underlying model is
made, which can be an expensive process. Further, SEOBNRv2 ROMs have only been generated for up
to four of the eight dimensions of intrinsic black-hole binary
waveform parameter space, for which only $\sim 10^7$ sequential
waveforms might need to be generated. To make matters
worse, due to technical challenges there is currently no method
available for efficiently producing ROMs to cover the full 
eight-dimensional precessing-spin parameter space.
Unless a breakthrough is made and
ROMs can be extended to all eight dimensions, their usefulness for
generic parameter estimation will be severely jeopardized (see
e.g.,~\cite{Farr2015}). At this point in time, the only way to generate
SEOBNR models over the full, eight-dimensional parameter space is to
use the (slow, not-fully-optimized) SEOBNRv3 code within {\tt LALSuite}
itself.

So, in anticipation of a breakthrough leading to SEOBNR ROMs over the
full eight-dimensional parameter space, and in preparation for the
release of the eight-dimensional SEOBNR approximant (SEOBNRv3, which
was undergoing code review while this paper was being written), in
this paper we describe our efforts in improving the performance of the
four-dimensional, second version of SEOBNR (SEOBNRv2) used in {\tt
LALSuite} by {\it more than two orders of magnitude} in terms of
execution time.
  
We stress that each optimization to SEOBNRv2 presented in this paper
has its analogue in SEOBNRv3, and some have already been included
in the SEOBNRv3 currently under development within {\tt LALSuite}.
Our fully optimized SEOBNRv2 now exists within {\tt LALSuite} under 
the moniker {\tt SEOBNRv2\_opt}.

After all of our optimizations, the overall speed-up factor of SEOBNRv2,
given by the waveform cycle-weighted average\footnote{over the three
  most promising sources of gravitational waves detectable by Advanced
  LIGO/VIRGO: double neutron star, black hole--neutron star,
  and black-hole binary systems},
\beq
\mathcal{S} = \frac{\sum_i \mathcal{S}_iN_{{\rm wc},i}}{\sum_i N_{{\rm
        wc},i}}\,, 
\label{eq:SpeedupFactor}
\eeq
was found to be $\sim 300$, where $\mathcal{S}_i$ is the speed-up
factor in generating the $i$'th waveform and $N_{{\rm wc},i}$ is the
number of wavecycles in the $i$'th waveform. 

Thus we have reduced the timescale for full PE using SEOBNRv2 {\it
directly} from 10--100 years to between {\it weeks} and {\it
months}, which now makes it possible to perform PE using SEOBNRv2 codes
{\it directly}. Core to our optimization philosophy was ensuring that
{\tt SEOBNRv2\_opt} agrees with the original SEOBNRv2 code to the
maximum degree possible (numerical roundoff error). To give an idea of
this very harsh standard we set for ourselves, of all 1,600 waveforms
sampled in this paper (across a wide swath of parameter space that
included all three compact binary systems of central interest to
Advanced LIGO/VIRGO),
the largest amplitude-weighted average phase difference (Eq.~\ref{eq:pherr})
was less than 0.008 rad (Table~\ref{tab:heat}). Advanced LIGO is
at best sensitive to phase differences of $\approx 0.17$ rad and
amplitude differences of order $10\%$, \cite{DetectionPaper2016} due to calibration
uncertainties. This is orders of magnitude beyond the maximum amplitude and phase
differences measured when comparing optimized versus unoptimized
SEOBNRv2 codes, so we conclude that our optimized SEOBNRv2 code can be used as a
drop-in replacement to the original code without any impact on PE.

\section{Optimization Strategies}
\label{OptStrats}
{\tt LALSuite}'s second SEOBNR code, SEOBNRv2~\cite{Taracchini2014}, which
models waveforms from aligned spin systems, is the focus of
optimization efforts in this paper.\footnote{An independent Spinning
EOB code that was recently developed \cite{Nagar2015}, but is not
currently available in {\tt LALSuite}, also addresses some of the
optimizations discussed in this paper, such as utilizing analytic 
derivatives over finite differences. They also claim significant 
speedups relative to the unoptimized version of SEOBNRv2 available 
in {\tt LALSuite}.} Next, we review the general strategies for 
optimizing SEOBNRv2, as well as the optimizations themselves.

Due to the necessity to frequently recompile and rerun the executables
during the optimization process, as well as the desire to isolate core
SEOBNRv2 functionality from other approximants within {\tt LALSuite},
we extracted the core functions required by the SEOBNRv2 approximant
from {\tt LALSuite} into an independent standalone code. This was
achieved through iteratively including functions the compiler
requested as necessary for compilation. Once completely extracted and
debugged, the standalone code's outputs were verified to be identical
to that of the original {\sc lalsimulation} software to within
roundoff error. 

\subsection{Basic Overview of SEOBNRv2 Code}
\label{v2overview}

For the purposes of optimization, we divide SEOBNRv2
into three components:
\begin{packed_enumerate}
\item {\bf ODE Solver}: The ``ODE Solver'' reads in initial
  parameters, generates initial data, and solves the SEOBNRv2
  Hamiltonian equations of motion \cite{Taracchini2012,Taracchini2014}.
  This part requires
  that several derivatives of the SEOBNRv2 Hamiltonian be evaluated
  $\gtrsim100,000$ times for a typical inspiral calculation starting
  at 10 Hz. Note that double neutron star (DNS) waveforms are the most
  costly to generate; for sources with a sufficiently large number of
  wave cycles, the overall time to generate a given waveform scales is
  dominated by this step, and scales roughly linearly with the number
  of cycles (as shown in the right panel of Fig.~\ref{fig:SU}).

\item {\bf Waveform Generation}: The full SEOBNRv2 time-domain
  waveforms are generally Fourier transformed into frequency space,
  which requires that the waveform be {\it evenly sampled} in
  time. However, for efficiency an {\it adaptive timestep}
  Runge-Kutta-type solver is adopted to solve the SEOBNRv2 equations
  of motion, so the ODE solution must be {\it interpolated} in time to
  achieve the evenly time-sampled solution required for the Fourier
  transform. As the input to this portion of the code is the solution
  to the SEOBNRv2 equations of motion and the output is the evenly
  time-sampled wave strain $h_{+,\times}(t)$, we call this the
  ``Waveform Generation'' part of the code.

\item {\bf QNM Attachment}: The SEOBNRv2 equations of motion will
  generate the full inspiral and merger portions of the waveform. In
  the ``QNM Attachment'' code, the quasi-normal mode (QNM) ringdown
  waveform is attached to the ODE solution. This component of the
  SEOBNRv2 code constituted an insignificant portion of the overall
  run time during the first $\sim$300x speed-up effort. After all the
  speed-ups were included, it contributed $\sim$20\% of the overall
  run time, so this portion of the code may need to be a focus
  of future optimizations.

\end{packed_enumerate}

\subsection{SEOBNRv2 Optimizations}
\label{v2opts}
We summarize the six highest-impact optimizations below, with timing
benchmarks ({\it italicized}) and speedup factors ({\it italicized, in
parentheses}) at each step of the optimization
process, using the 16,277 wave cycle $1.4\sunmass+1.4\sunmass$ double
neutron star binary merger scenario as our test for benchmarking (with
a start frequency of 10 Hz). Double neutron star binaries were chosen to measure
overall speed-up factors because they are the most costly to
generate.

\begin{packed_enumerate}
\item[0.] {\it 370 s} Original, un-optimized SEOBNRv2.

\item {\it 197 s (1.9x)} Change from the \verb|gcc|~\cite{gcc} to
the \verb|Intel| compiler v15.0.1~\cite{icc}. The Intel Compiler
Suite (\verb|Intel| compiler) is well known to often generate far more
optimized executables than the GNU Compiler Collection (\verb|gcc|).

\item {\it 133 s (2.8x)} Hand-optimize SEOBNRv2 Hamiltonian algebraic
expressions, removing all unnecessary calls to expensive
transcendental functions,  like \verb|exp()| and \verb|log()|.

\item {\it 23.1 s (16x)} Replace all finite difference derivatives of the
SEOBNRv2 Hamiltonian with exact expressions automatically generated by
Mathematica~\cite{mathematica}.

This replacement improves code performance in
two ways. First, it reduces the large amount of roundoff-error-induced
noise from finite differencing, enabling the adaptive-timestep ODE
solver (Runge-Kutta fourth order) to achieve the desired accuracy in fewer
steps. Second, evaluating derivatives exactly calls for significantly
fewer arithmetic operations than calling the original Hamiltonian
multiple times as required to compute a finite difference derivative.

\item {\it 7.06 s (52x)} Reduce number and cost of interpolations inside
the Waveform Generation component of the code. 

After the adaptive-timestep Runge-Kutta fourth order ODE Solver has
completed, the solution to the equations of motion is sparsely and
unevenly sampled in time. However, the final waveform needs to be uniformly
sampled in time, as it will be fed to a fast Fourier transform
(FFT) algorithm (outside of SEOBNRv2). The solution is therefore
interpolated (inside of SEOBNRv2) to the desired {\it constant}
sampling frequency using cubic splines.

The original Waveform Generation code interpolated the four evolved 
variables, $r$, $p_r$, $\phi$, and $p_{\phi}$ \footnote{For further
discussion of the equations solved in SEOBNRv2, see Section 2
of~\cite{Taracchini2014}.}, then reconstructed the waveform amplitude
and phase from these variables at each interpolated point. We modified
the code to compute amplitude and phase from the sparsely-sampled ODE
solution first, and then interpolate amplitude and phase directly at
the desired points (as these change as rapidly in time as the evolved
variables). While reducing the number of amplitude and phase
calculations greatly improved performance, there still exists
dependencies on the four evolved  variables, necessitating that they
still be interpolated as well. This will be the focus of future
optimization efforts. 

Also, the cubic spline interpolation routine, which is built-in
to the GNU Scientific Library (GSL)~\cite{gsl}, recomputes interpolation coefficients
at each point. Since we wish to interpolate the very sparsely-sampled ODE
solution to a fixed frequency, these interpolation coefficients are
often recomputed thousands of times over. We therefore inserted the GSL
spline functions into our optimized SEOBNRv2 code and modified them so
that these coefficient calculations are not unnecessarily repeated.

\item {\it 2.81 s (130x)} Remove unnecessary recalculations within the
  ODE Solver. Inside the ODE Solver is a loop over $\ell$ and $m$
  modes, within which the orbital angular velocity $\omega$ is
  computed for each $\ell$ and $m$. Computing $\omega$ is a
  particularly expensive operation, requiring one evaluation of an
  SEOBNRv2 Hamiltonian derivative. However, $\omega$ does not depend
  on $\ell$ and $m$, so it only needs to be computed once and not for
  all $\ell$ and $m$.

\item {\it 1.53 s (240x)} Increasing order of ODE Solver to Runge
  Kutta eighth order (RK8)~\cite{gsl}. After replacing all finite
  difference derivatives within the ODE Solver with analytical
  derivatives, the ODE Solver required far fewer steps to achieve the
  desired precision. Encouraged by this, we experimented with
  different ODE integrators, and found RK8 to be far more efficient
  than RK4~\cite{gsl}, particularly in the case of long waveforms.
\end{packed_enumerate}

\section{Results}

\subsection{Performance Benchmarks}

To demonstrate how much we were able to speed-up the different
components of the SEOBNRv2 code for scenarios of key interest to
Advanced LIGO/VIRGO, Table~\ref{midbench} presents timing
results for both the original and fully-optimized SEOBNRv2 codes,
splitting the timings so that speed-up factors of individual code
components could be exposed. Notice that the
``Waveform Generation''+``QNM Attachment'' components were optimized
uniformly by a factor of $\sim 400$x, regardless of the length of the
waveform. This most significantly impacts the performance of physical
scenarios in which these components dominated the runtime, most
notably scenarios that spend the shortest times in-band, like black
hole binaries (BHBs). The {\it total}
speed-up factor is dominated by the ``ODE Solver'' speed-up, which
gradually increases with the number of wavecycles, from 
$\sim40$x for $\sim 600$ wavecycle BHBs, to $\sim 100$x for 
$\sim 4,000$ wavecycle scenarios (black hole--neutron star binaries,
BHNS), to $\sim 210$x for $\sim 16,000$ cycle double neutron star
(DNS) waveforms.

Next, we surveyed the likely four-dimensional parameter space of
binaries observable by Advanced LIGO/VIRGO, as specified in
Table~\ref{tab:ranges}. In short, we generated a total of 400
waveforms for each of four scenarios, all with a start frequency of
10Hz. Each scenario samples two dimensions of parameter space,
choosing 20 values for parameters $Q_1$ and $Q_2$. The first scenario
considered a black-hole binary system (BHB$_{\rm M}$) in which total
mass and mass ratios were varied. The second also considered a BHB
system, but with equal masses and varied spins (BHB$_{\rm S}$). The
third varied the mass and spin of the black hole in a black
hole--neutron star binary (BHNS) system, with a 1.4$\sunmass$ neutron
star. The fourth scenario was that of a spinless double neutron star
system (DNS) with variable masses.

\begin{figure}[!h]
\subcaptionbox{\label{fig:SUvWC}}
  {\includegraphics[width=2.95in]{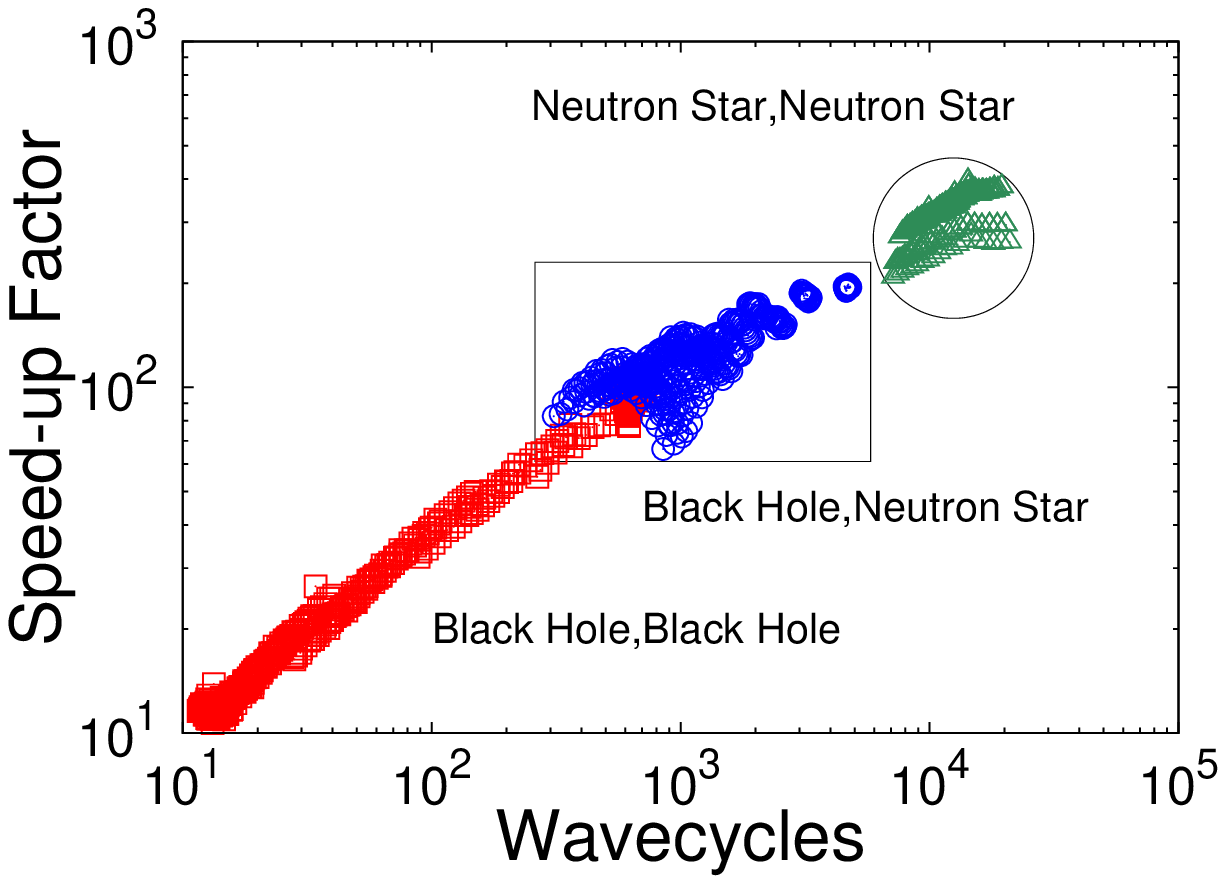}}\hfill
\subcaptionbox{\label{fig:TvWC}}
  {\includegraphics[width=2.95in]{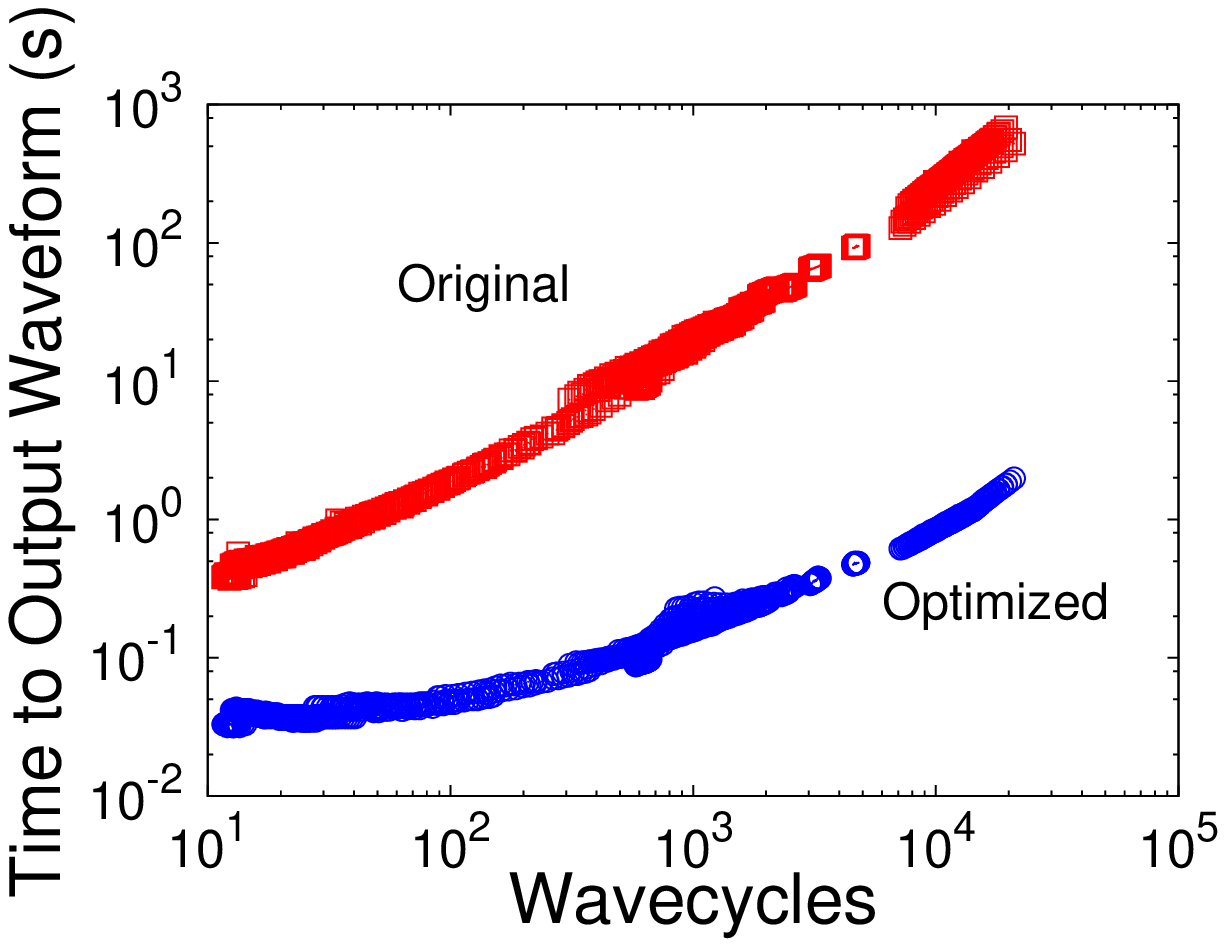}}
  \caption{\label{fig:SU} {\bf Performance Benchmarks}: {\bf Left panel}:
    Speed-up factor for (fully) Optimized versus Original SEOBNRv2,
    considering key compact binary systems of interest to Advanced
    LIGO. {\bf Right panel}: Total time required to produce SEOBNRv2
    waveforms for the same compact binary systems (in seconds). The
    ratio of the top curve to the bottom curve in the right panel for
    a given scenario is equivalent to the speed-up factor of the left
    panel. Initial parameters for waveforms are as specified in
    Table~\ref{tab:ranges}, with start frequency of 10 Hz and
    output sample rate set to 16,384 Hz. Performance was
    measured on a computer with an Intel i7-4790 CPU and 16 GB of
    RAM.}
\end{figure}

As we generated waveforms in this survey, we performed both
performance benchmarks and error analyses, of both the original
(un-optimized) and fully-optimized SEOBNRv2 codes. As shown in the left
panel of Fig.~\ref{fig:SU}, we immediately find that the pattern first
observed in Table~\ref{midbench} is a general one: as the number of
wavecycles increases, the speed-up factor increases
significantly. Most importantly, the right panel of Fig.~\ref{fig:SU}
indicates that with modern CPUs, we can now generate practically {\it
  all} SEOBNRv2 waveforms of interest to Advanced LIGO/VIRGO in under
two seconds, whereas with the original code without optimizations, the most expensive
waveforms require roughly 12 minutes to generate, representing a
speed-up factor of roughly 400 for these particularly difficult cases.

\begin{figure}[!h]
$
\begin{aligned}
\begin{subfigure}{.3\linewidth}
\includegraphics[,width=.97\linewidth]{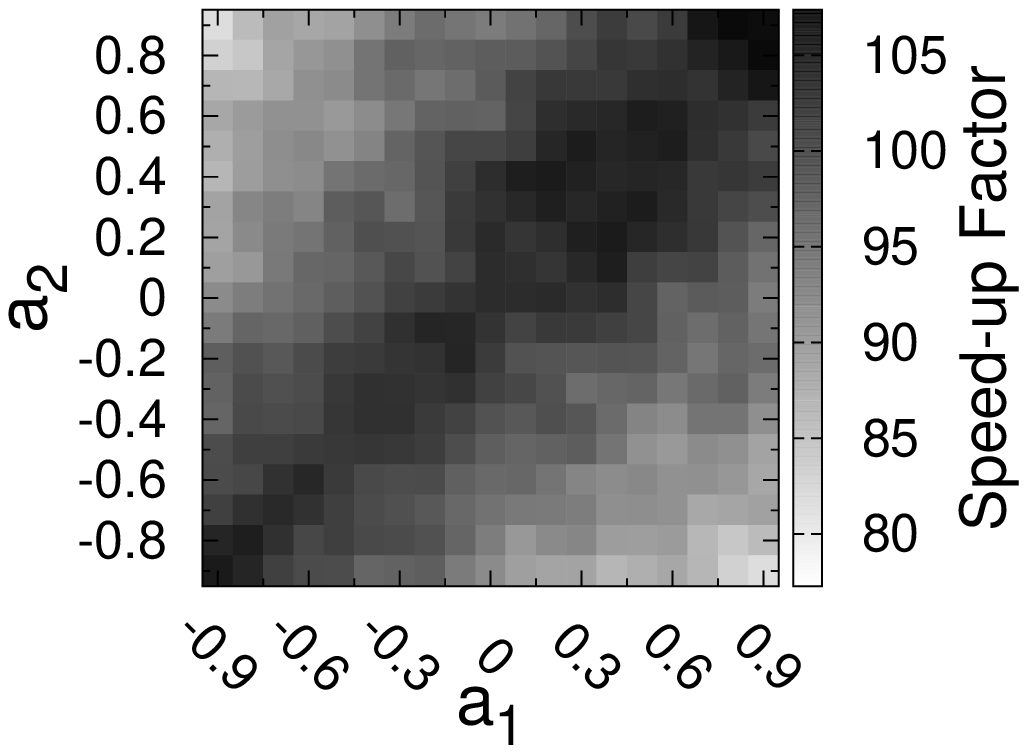} \caption{\label{fig:SU1}}
\end{subfigure}
\hspace{.15cm}
\begin{subfigure}{.3\linewidth}
\includegraphics[,width=.97\linewidth]{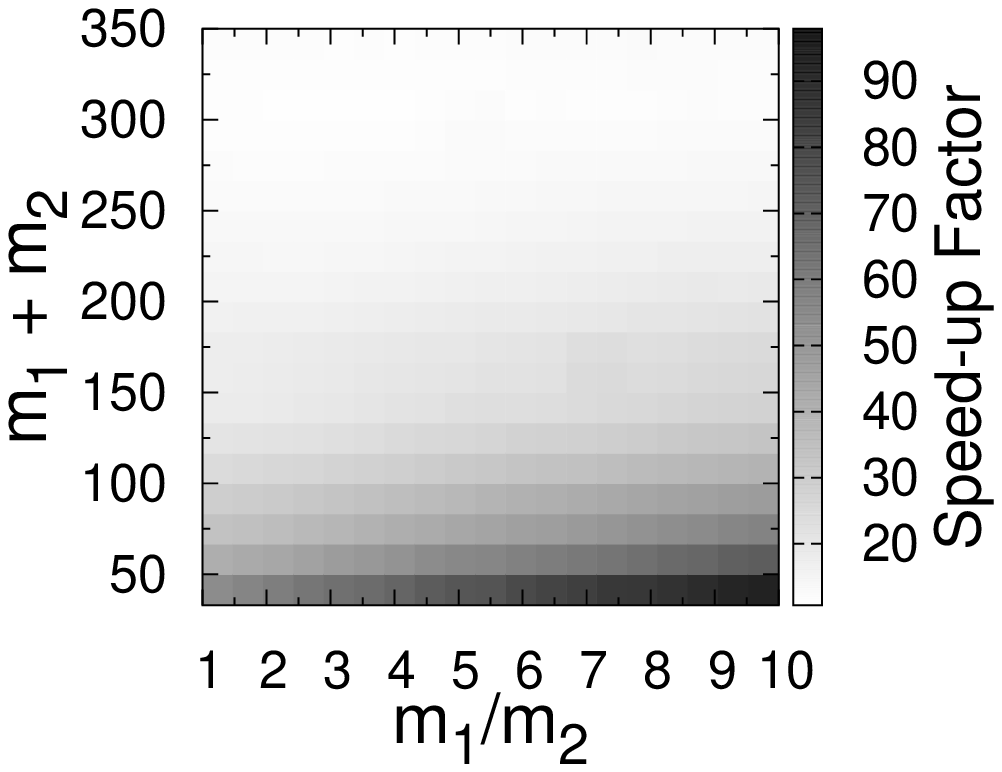} \caption{\label{fig:SU2}}
\end{subfigure}\\
\begin{subfigure}{.3\linewidth}
\includegraphics[,width=.97\linewidth]{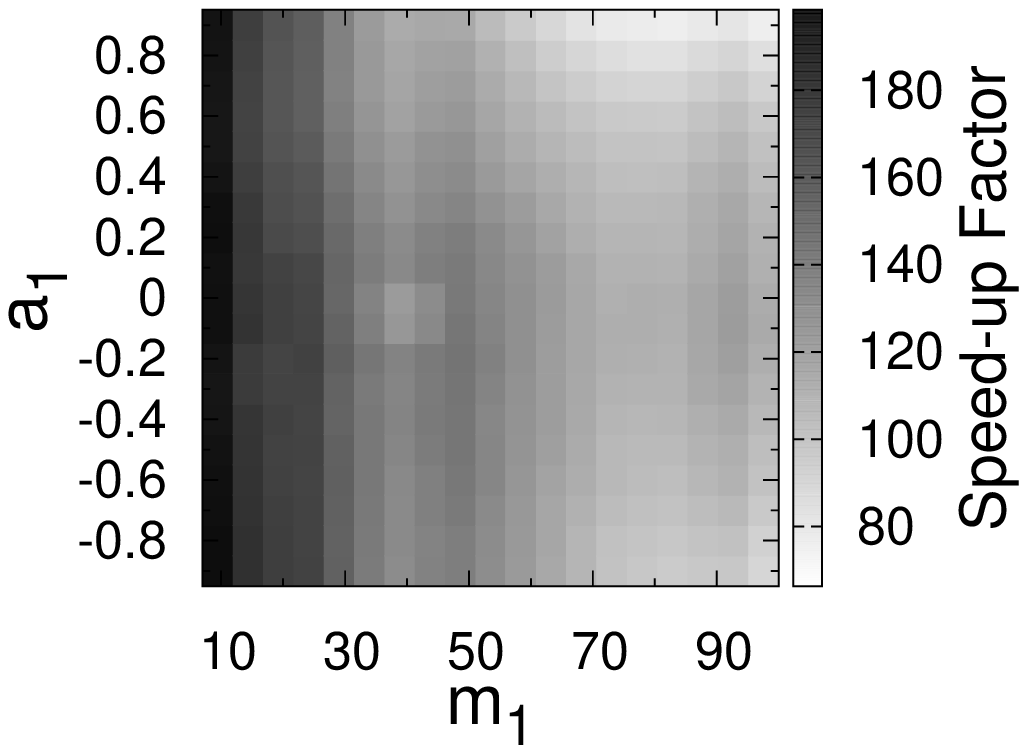} \caption{\label{fig:SU3}}
\end{subfigure}
\hspace{.15cm}
\begin{subfigure}{.3\linewidth}
\includegraphics[,width=.97\linewidth]{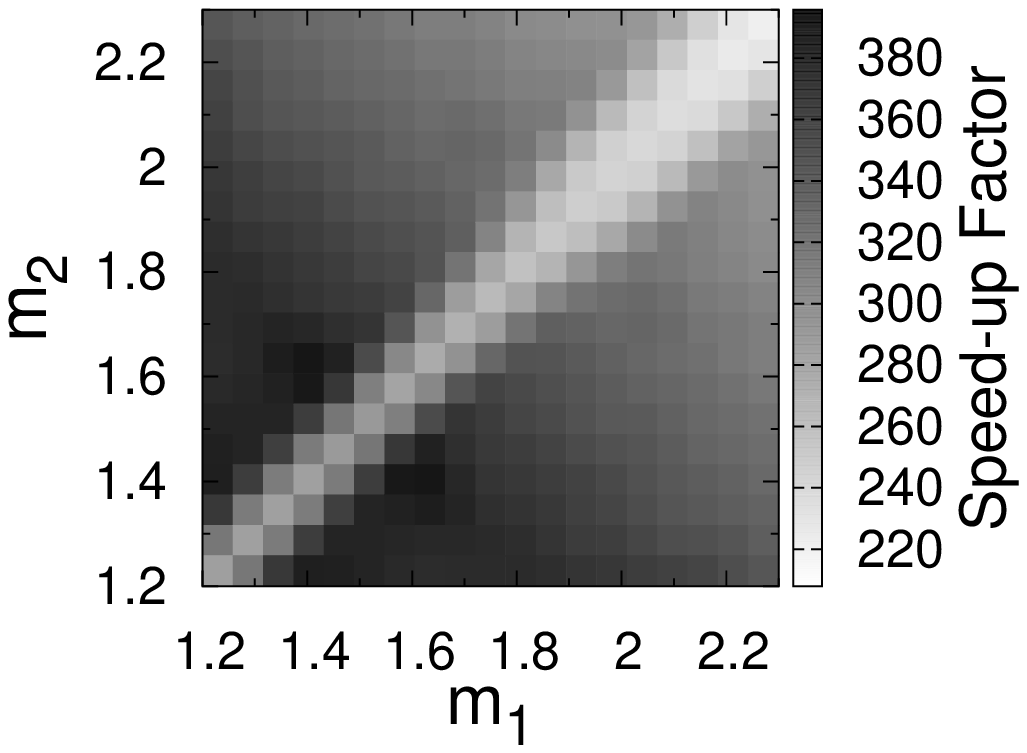} \caption{\label{fig:SU4}}
\end{subfigure}
\end{aligned}
$
\caption{\label{fig:SU0} {\bf Speed-ups over parameter spaces:}
  Measured speed-up factors across SEOBNRv2 parameter space, where
  speed-up factor is defined as the ratio of run-time of the original
  SEOBNRv2 code to our optimized version. Note that these results
  use exactly the same data as in Fig.~\ref{fig:SU}. The shading of each of the
  1,600 pixels in these panels corresponds to the measured speed-up
  factor for one of the 1,600 cases specified in
  Table~\ref{tab:ranges}, with darker shading corresponding to larger
  speed-ups. Each panel corresponds to a different physical scenario,
  with the {\bf top-left} denoting BHB$_{\rm s}$, {\bf top-right}
  BHB$_{\rm m}$, {\bf bottom-left} BHNS, and {\bf bottom-right} DNS.
}
\end{figure}

For finer-grained analysis, we also visualize the speed-ups plotted in
Fig.~\ref{fig:SU} using ``heat maps'' in Fig.~\ref{fig:SU0}. In
addition to the pattern observed in Fig.~\ref{fig:SU}, by which the
speed-up factor is found to increase with the number of wave cycles
(corresponding to a decrease in total mass in Panels~\ref{fig:SU2},
\ref{fig:SU3}, and \ref{fig:SU4}), we also find that the speed-up
factor decreases as the mass ratio approaches unity
(Panel~\ref{fig:SU4}), and when the ratio of the spin 
magnitudes for BHBs deviates from unity (Panel~\ref{fig:SU1}).

Interestingly, the clear feature along the positive
diagonal of Panel~\ref{fig:SU4} is not consistent with the pattern of
increasing speed-up factor with increasing number of wavecycles.
It turns out that this feature spawns from branchings within the
SEOBNRv2 code to handle equal mass cases. While this feature and the
small quantitative variations in the ``heat maps'' of Fig.~\ref{fig:SU0}
are consistent with the fact that the pattern in the left panel of Fig.~\ref{fig:SU}
exhibits scatter from an otherwise clear power-law trend, the results
of Fig.~\ref{fig:SU} are convincing enough to lead us to conclude that
these minor patterns observed in mass ratio and spin parameter ratio
have little impact on the speed-up factor, as compared to the number
of wavecycles.

The rightmost column of Table~\ref{tab:heat} demonstrates that our
average speed-up factor for BHBs ranges between
$\sim50$--100, increasing to 145 for BHNS, and 322 for DNS, again
consistent with the leading pattern that speed-up factors increase
with the number of wavecycles in band.

\begin{table}[!htp]
\begin{center}
\begin{tabular}{|p{1.35 in}|p{0.75in}|p{0.75in}|p{1.15in}||p{0.6in}|}
\hline
Physical & Code & ODE & Waveform Gen. & Total \\
 Scenario &  & Solver & \& QNM Attach. & \\
\hline
BHB, spins=0 & Original & 3.129s & 6.247s & 9.376s  \\
($10\sunmass$+$10\sunmass$) & Optimized & 0.076s & 0.015s & 0.091s \\
Wavecycles: 621 & & x({\bf41.17}) & x({\bf 416.5}) & x({\bf 103.0}) \\
\hline
BHB, spins=0.9 & Original & 3.659s & 6.595s & 10.254s  \\
($10\sunmass$+$10\sunmass$) & Optimized & 0.081s & 0.016s & 0.097s \\
Wavecycles: 664 &  & x({\bf45.17}) & x({\bf412.2}) & x({\bf105.7}) \\
\hline
BHNS, spins=0 & Original & 39.877s & 36.733s & 76.610s  \\
($1.4\sunmass$+$10\sunmass$) & Optimized & 0.310s & 0.083s & 0.393s \\
Wavecycles: 3609 & & x({\bf128.6}) & x({\bf442.6}) & x({\bf194.9}) \\
\hline
BHNS, $s^z_{\rm BH}=0.9$ & Original & 42.727s & 38.292s & 81.019s  \\
($1.4\sunmass$+$10\sunmass$) & Optimized & 0.327s & 0.079s & 0.406s \\
Wavecycles: 3753 & & x({\bf130.7}) & x({\bf484.7}) & x({\bf199.6}) \\
\hline
DNS, spins=0 & Original & 209.146s & 161.111s & 370.257s  \\
($1.4\sunmass$+$1.4\sunmass$) & Optimized & 0.965s & 0.438s & 1.403s \\
Wavecycles: 16277 & & x({\bf216.7}) & x({\bf367.8}) & x({\bf263.9}) \\
\hline
\end{tabular}
\end{center}
\caption{{\bf Timing benchmarks, selected cases of interest:}
  Comparison between the original SEOBNRv2 code with the fully
  optimized version. A starting frequency of 10 Hz was chosen for each
  scenario, with a waveform sample rate of 16,384 Hz. These waveforms
  were generated on a modern computer, with an Intel i7-4790 CPU and
  16 GB of RAM.}
\label{midbench}
\end{table}

\begin{table} [!h]
  \begin{center}
    \begin{tabular}{l|cccc}
      ranges & $m_1$ $(\sunmass)$ & $q$ & $a_1$ & $a_2$\\\hline
      ${\rm BHB_M}$ & $Q_1\in[16.7\dots309]$ & $Q_2\in[1... 10]$ & $0$ & $0$\\
      ${\rm BHB_S}$ & $10$ & $1 $ & $Q_1\in[-0.95...0.95]$ & $Q_2\in[-0.95...0.95]$ \\      
      ${\rm BHNS}$ & $Q_1\in[7...100]$ & $\frac{m_1}{1.4 \sunmass}$ & $Q_2\in[-0.95...0.95]$ & $0$\\
      ${\rm DNS}$ & $Q_1\in[1.2...2.3]$ & $\frac{m_1}{Q_2\in[1.2...2.3\sunmass]}$ & $0$ & $0$\\
    \end{tabular}
  \caption{\label{tab:ranges}{\bf Surveyed parameters}: We survey a
    total of four physical scenarios: a BHB mass survey
    (BHB$_M$), BHB spin survey (BHB$_S$), BHNS BH mass and spin survey
    (BHNS), and a DNS mass survey (DNS). $m_1$, $m_2$ denote the
    compact objects' masses; $q$ is the mass ratio ($m_1/m_2$); and
    $a_1$,$a_2$ are the corresponding dimensionless Kerr spins for
    each compact object. Each calculation was performed from a
    start frequency of 10 Hz. $Q_1$ and $Q_2$ are the variable
    parameters in each physical scenario, taking on 20 values in each
    of the indicated ranges.}
  \end{center}
\end{table}

\begin{table}[h]
\begin{tabular}{|l|ccc|ccc|c|}
\hline
\multirow{2}{*}{\parbox{2cm}{Physical Scenario}} & \multicolumn{6}{|c|}{Error Maxima} & Average Speed-up\\\hhline{~-------}
& $||\Delta_{\phi}||_A$ &  $Q_1$ & $Q_2$ & $||\Delta_\mathcal{A}||_A$ & $Q_1$ & $Q_2$ & $\mathcal{S}$ \\ \hline
${\rm BHB_S}$ &0.00735 & -0.850 & -0.750 & -4.01 & -0.650 & 0.050 & 96.4\\ 
${\rm BHB_M}$&0.00523 & 4.32 & 167. & -3.81 & 1.94 & 333. & 53.7 \\
${\rm BHNS}$&0.00790 & 41.3 & 0.950 & -4.30 & 26.6 & 0.750 & 145\\
${\rm DNS}$&0.00465 & 1.66 & 1.66 & -4.09 & 1.20 & 1.49 & 322\\ \hline
\end{tabular}
\caption{\label{tab:heat} {\bf Maximum Errors and Overall Speed-ups:}
  Summary of results of parameter survey (as specified in
  Table~\ref{tab:ranges}). $||\Delta_{\phi}||_A$ and
  $||\Delta_\mathcal{A}||_A$ denote the maximum amplitude weighted
  amplitude and phase errors (Eqs.~\ref{eq:amperr} and
  \ref{eq:pherr}). $Q_1$ and $Q_2$ (See Table \ref{tab:ranges}) are
  the variable parameters resulting in the maximum
  $||\Delta_{\phi}||_A$ and $||\Delta_\mathcal{A}||_A$. $\mathcal{S}$
  is the weighted average speed-up for each scenario (Eq.~\ref{eq:SpeedupFactor}).}
\end{table}

\subsection{Code Validation Tests}

We conclude that our performance speed-ups are substantial, but unless
we can {\it guarantee} that the optimized code generates waveforms
that agree as well as numerically possible with the original (i.e., to
roundoff-error), then we could not be confident that we did not
introduce some error in the code. Defining quantities most useful for
such error analysis was a difficult task, as with the original
SEOBNRv2 code, perturbations at the 15th significant digit in initial
parameters could, e.g., generate amplitude differences of order unity
at the end of the ringdown, where the amplitude was of order $10^{-6}$
the maximum amplitude, yet at all other points the amplitudes would
agree to many ($\sim6+$) significant digits. To mitigate this effect,
we compute the amplitude-weighted relative amplitude error:
\beq
\label{eq:amperr}
||\Delta A||_A = \frac{\sum_t A_t\log_{10}(|{A_1}_t-{A_2}_t|/|{A_1}_t|)}{\sum_t |A_t|}\,,
\eeq
where ${A_1}_t$ and ${A_2}_t$ are the two compared amplitudes at time
$t$ with ${A_1}_t$ taken to be the larger of the two amplitudes and
${A_t}$ chosen to be the amplitude at time $t$ of one waveform,
consistently. Similarly for phase, the amplitude-weighted phase error
was computed,
\beq
\label{eq:pherr}
||\Delta \theta||_A = \frac{\sum_t |A_t({\theta_1}_t - {\theta_2}_t)|}{\sum_t |A_t|}
\eeq
where ${\theta_1}_t$ and ${\theta_2}_t$ were the compared phases, and
$A_t$ is the same as in the amplitude error calculation.

The algorithm to compare two waveforms works as follows. The waveforms
output by {\tt LALSuite} are output with uniform timesteps such that
the time of peak amplitude is chosen to be $t=0$ seconds.
If the initial parameters are perturbed infinitesimally (i.e., at the
15th significant digit), or even if different compilers/compiler flags
are used to generate the trusted, un-optimized SEOBNRv2 executable,
then this peak shifts slightly, shifting the position of t=0 slightly;
this sometimes changes the number of data points prior to the peak.
As a result, an
additional point or two may be added to the start of the waveform with the
later peak. Let's call this waveform A. Naturally, the tail of
waveform A will generally truncate a couple of points earlier than the
other waveform (waveform B). 

So the first step in our error analysis removes data points from the
start of waveform A until the initial time is within a single
timestep of the start of waveform B. The same number of data
points are removed from the tail of waveform B.

Since an overall small time shift may still be present between the
waveforms, the amplitudes and phases of one waveform are quadratically
interpolated to the moments in time provided by the other
waveform. Once the amplitude of either waveform dropped to zero, the
times of the last two nonzero amplitudes and all times which followed
were truncated. 
 Finally, the amplitude and phase errors
(Eqs.~\ref{eq:amperr} and \ref{eq:pherr}) were computed.

\begin{figure}[!h]
\begin{subfigure}{.47\linewidth}
\includegraphics[width=.97\linewidth]{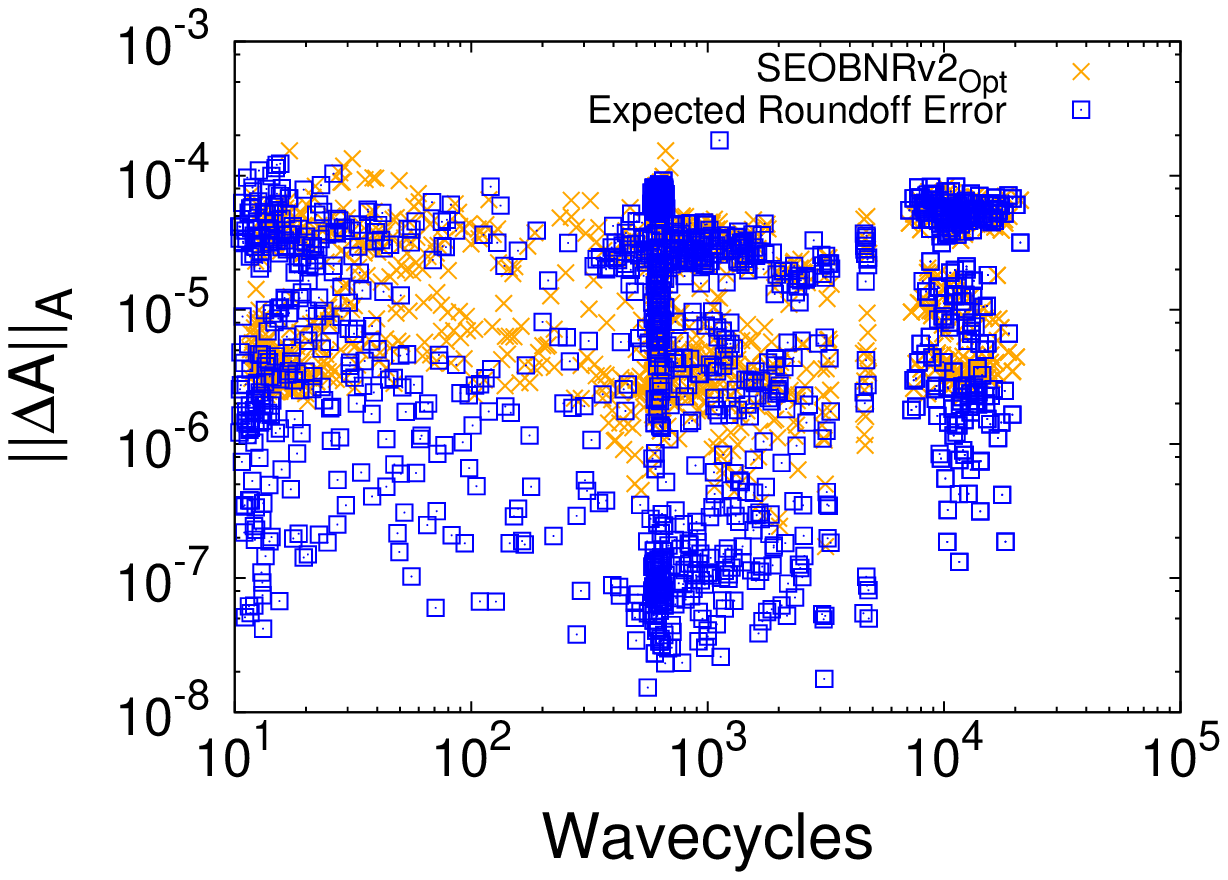} \caption{\label{fig:RoundOffALAL}}
\end{subfigure}\hspace{.15cm}
\begin{subfigure}{.47\linewidth}
\includegraphics[width=.97\linewidth]{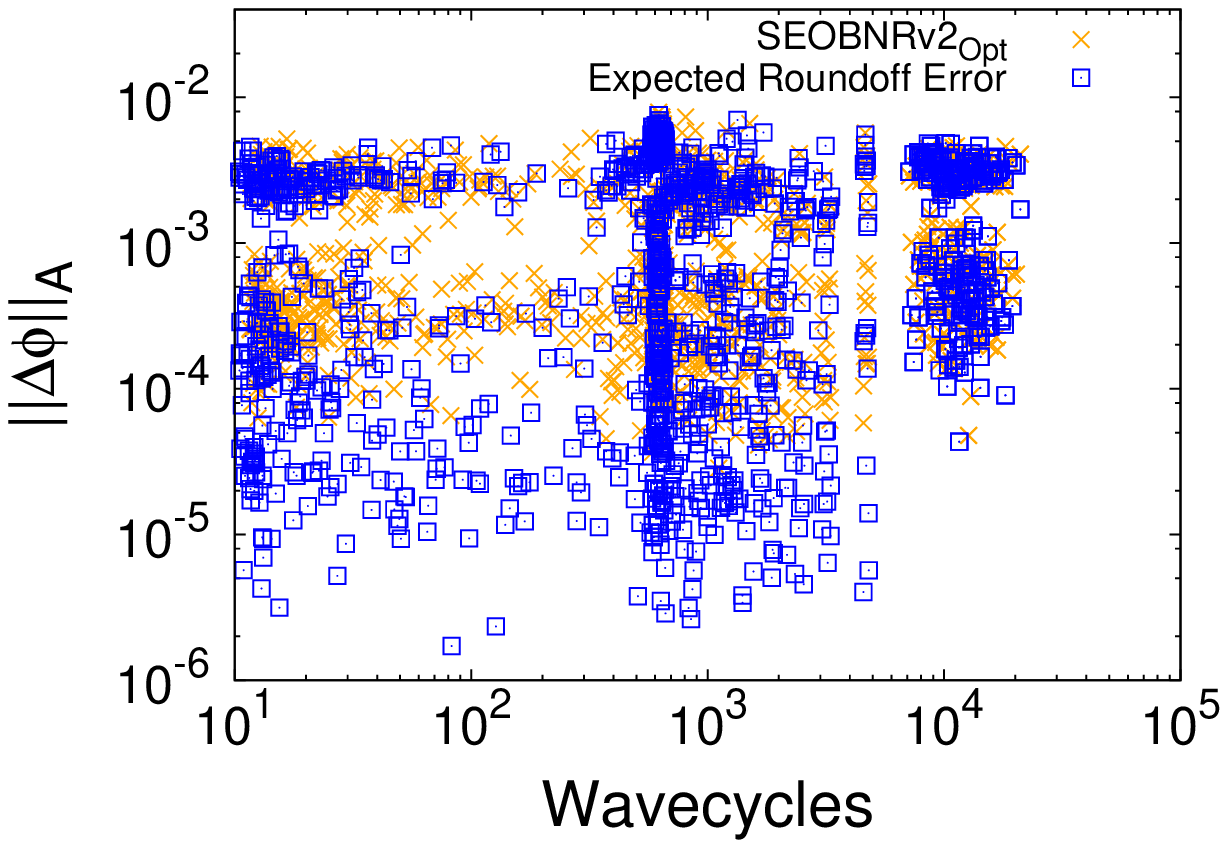} \caption{\label{fig:RoundOffPLAL}}
\end{subfigure}\hspace{.15cm}

\caption{\label{fig:RoundOffLAL} {\bf roundoff Error vs Optimization
    Mismatch:} The amplitude {\bf left} and phase {\bf right} errors
  vs wavecycles, compared to the error caused by an
  order-machine-epsilon perturbation in one of the masses (Roundoff
  Error), for all of the surveyed parameter space. Both panels
  demonstrate much overlap between these quantities.} 
\end{figure}

Figure~\ref{fig:RoundOffLAL} shows that the phase and amplitude
differences between {\tt LALSuite}'s SEOBNRv2 and our {\tt
  SEOBNRv2\_opt} (Eqs.~\ref{eq:amperr} and \ref{eq:pherr}) are
stochastically distributed across our 1,600-point parameter
survey. To measure the magnitude of roundoff errors in the original
code, we computed the same differences, comparing instead the original
code with itself but with a small, 15th-significant-digit perturbation
to the input parameters (e.g., $m_1\to m_1(1+10^{-15})$). I.e., we
measure the expected roundoff errors by perturbing the input
parameters by a small multiple of double-precision machine epsilon.

The overlap between the expected roundoff error and the error
in the optimized code is striking, consistent with the explanation that
differences between the optimized and un-optimized codes are
completely due to roundoff error. 

\begin{figure}
  \begin{center}
    \includegraphics[,width=.49\linewidth]{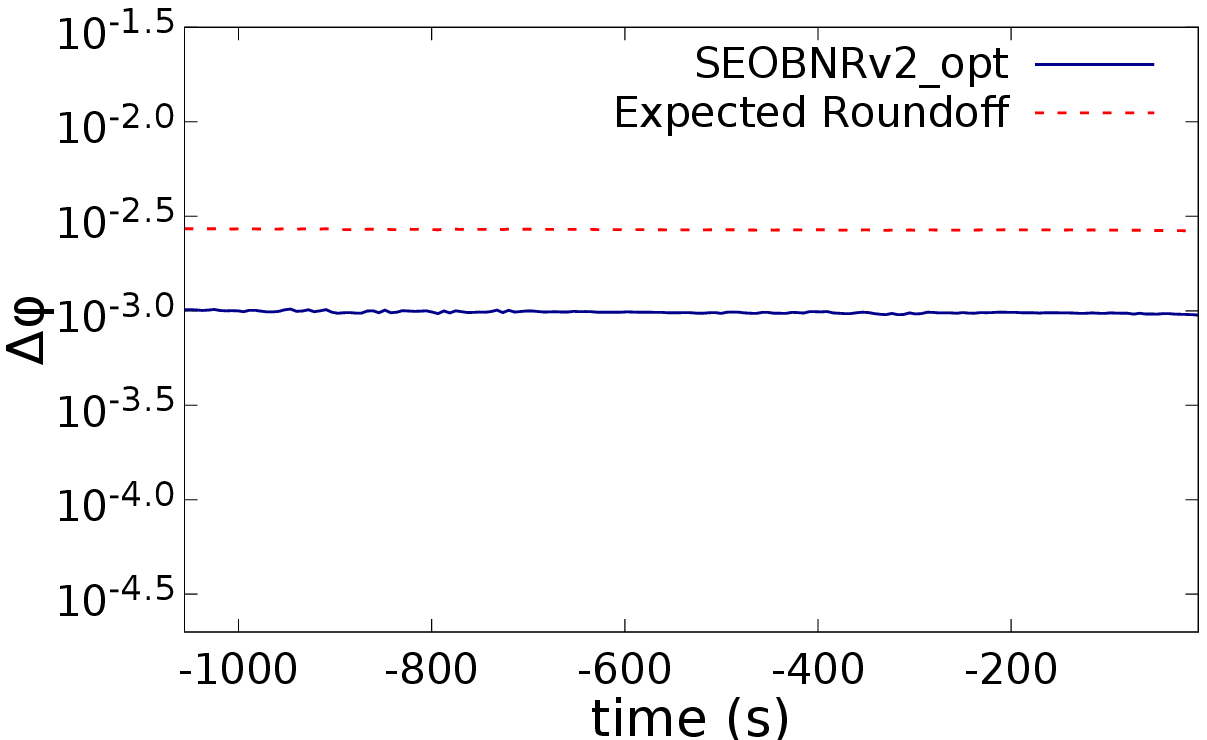}\hfill
    \includegraphics[,width=.49\linewidth]{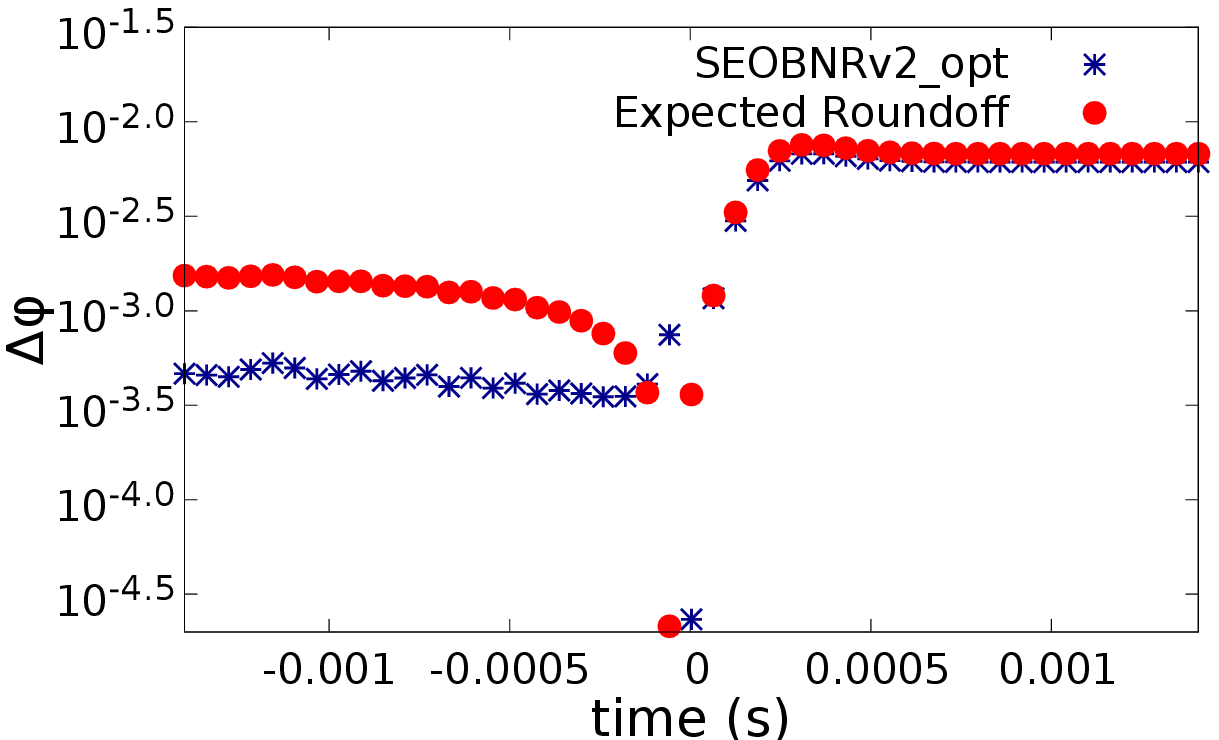}
    \caption{\label{fig:phasevst} \textbf{Phase Differences vs Time:}
      Shown is a typical example of the phase differences ($\Delta
      \phi$) near merger (\textbf{right panel}) and before
      (\textbf{left panel}), between a waveform generated by SEOBNRv2
      and a waveform generated by {\tt SEOBNRv2\_opt} using the same input
      parameters, as well as the phase differences between the same
      waveform generated by SEOBNRv2 code and that generated by
      SEOBNRv2 with input parameters were perturbed to the order of
      $10^{-15}$. This value is taken to be the expected roundoff
      error. This system corresponds to a nonspinning DNS system with masses $1.258$
      and $1.489$. Notice that the phase differences are
      roughly constant until near merger.
}
  \end{center}
\end{figure}

We focused our efforts on computing amplitude and phase errors in the
region near merger, because, as shown in Fig.~\ref{fig:phasevst} this
region contains the largest phase errors. Fig.~\ref{fig:phasevst} also
demonstrates that no significant phase difference had accrued up to
this region. Regardless, to be absolutely certain in our results, we
repeated our full amplitude and phase discrepancy analysis on data
using only the first 10,000 output times of each waveform as well, and
precisely the same degree of overlap between optimized and
un-optimized waveform amplitude and phases was observed.

We summarize in Table~\ref{tab:heat} the cases with the
largest discrepancies in amplitude and phase between the optimized and
un-optimized codes, for each of the four scenarios. Notice that there
is no clear pattern in worst-case parameters, consistent with the
stochastic nature of phase and amplitude errors observed in
Fig.~\ref{fig:RoundOffLAL}. In addition, we observe in the worst case
discrepancies in phase of roughly 0.008 rad and relative amplitude of
$10^{-3.8}$.

\subsection{Improved Sensitivity to Initial Parameters}

The SEOBNR codes are designed to read in start frequency and intrinsic
binary parameters such as initial masses and spins. The
codes then use these parameters to generate initial conditions for radius
$r$, radial momentum $p_r$, and $\phi$-component of momentum $p_\phi$
for the system, which are required inputs to the SEOBNR Hamiltonian
equations of motion. In computing Hamiltonian input
quantities consistent with the initial parameters, the SEOBNR
codes evaluate finite difference derivatives of the Hamiltonian.

Unfortunately, it is well known that finite difference derivatives can
be prone to enormous roundoff errors, which act to enormously amplify
small perturbations in chosen initial binary parameters, yielding an
observable effect on amplitude and phase of resulting
waveforms. For example, if the initial mass of a single binary
component is perturbed by anywhere from one part in $10^{15}$ to
one part in $10^8$, the resulting perturbation in $p_\phi$ input into
the Hamiltonian will be stochastic, with perturbation amplitude fixed
at one part in $10^9$, thus making SEOBNRv2 completely incapable of
exploring initial mass perturbations beyond the eighth significant
digit.

\begin{figure}[H]
  \subcaptionbox{}
  {\includegraphics[width=1.95in]{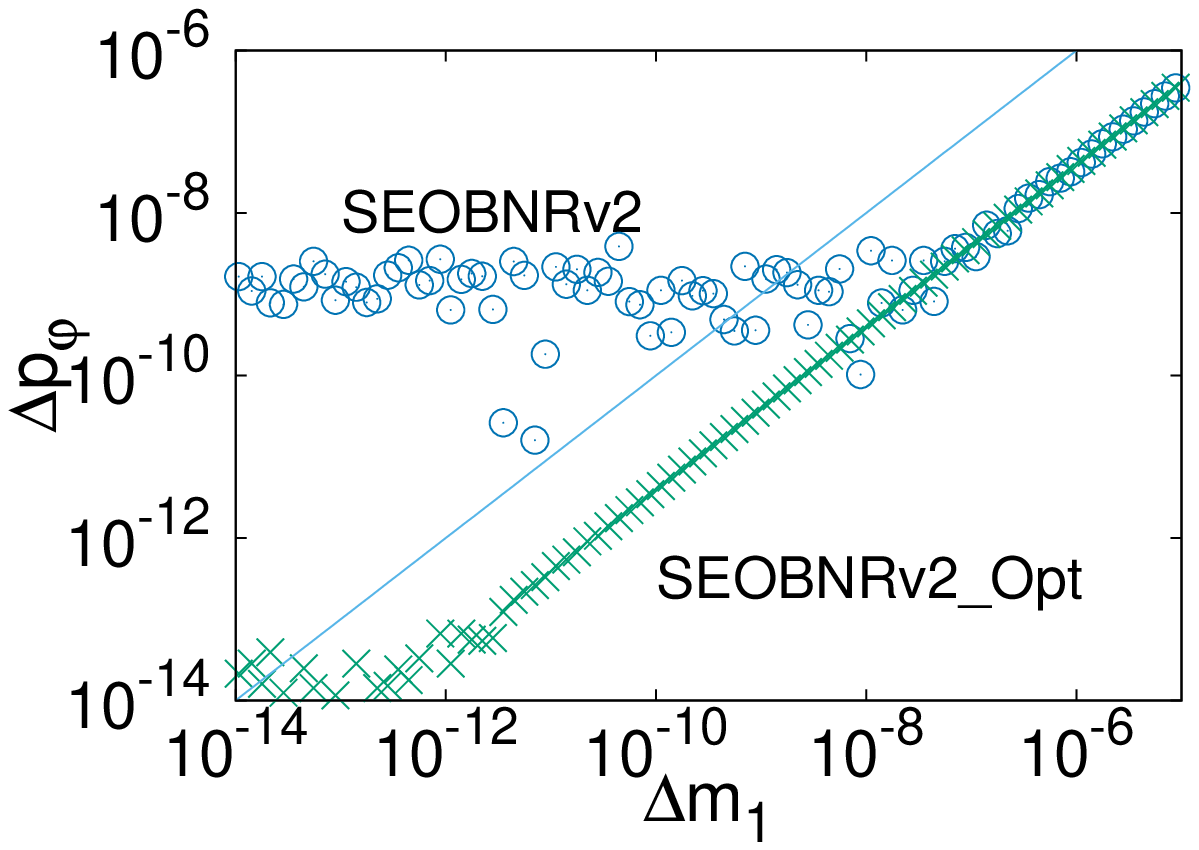}}\hfill
  \subcaptionbox{}
  {\includegraphics[width=1.95in]{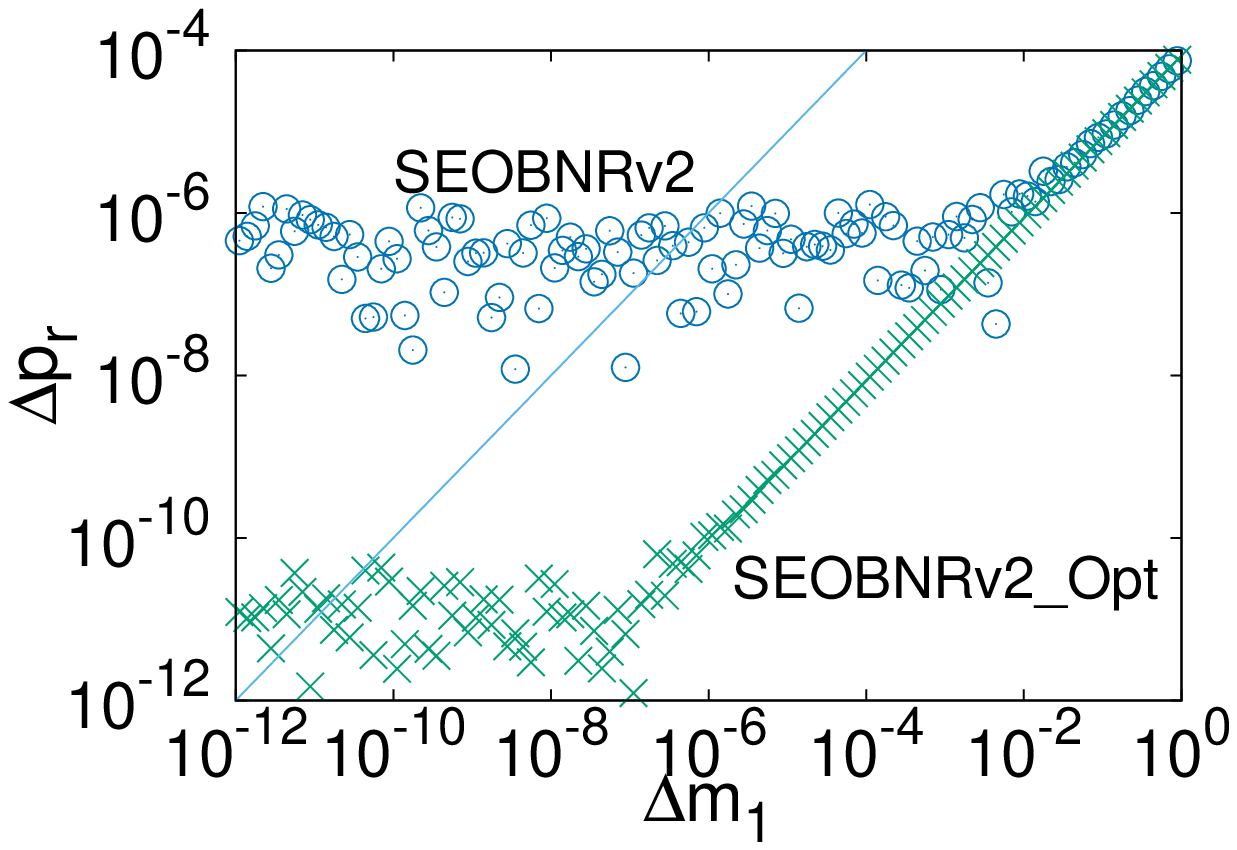}}\hfill
  \subcaptionbox{}
  {\includegraphics[width=1.95in]{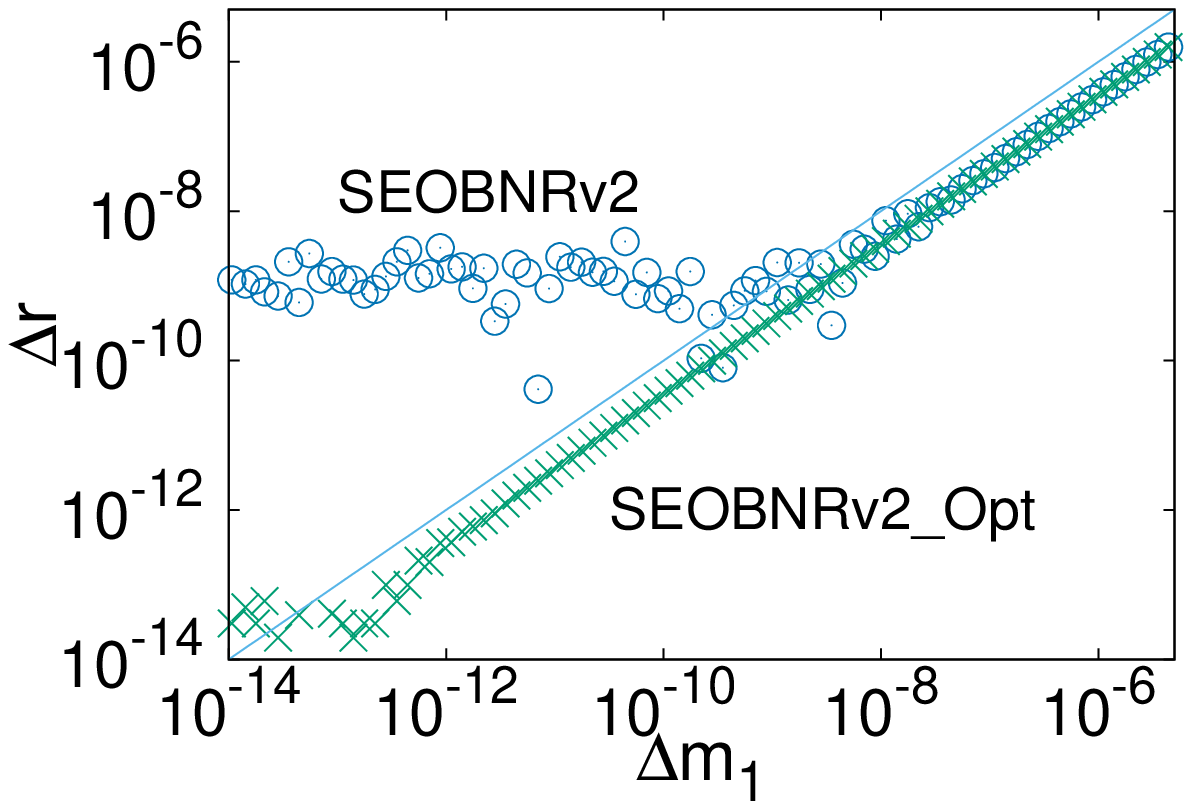}}
  \caption{\label{fig:Snstvtyeps} {\bf Initial data sensitivity mass
      perturbation:} The dependence of the initial values of $p_r$ and
    $p_\phi$ on the mass, $m_1$ of one compact object of the system,
    in both the original code and the optimized version.
      $\Delta p_{\phi}$,$\Delta p_{r}$, and
    $\Delta r$ are the differences in the initial values of
    $p_{\phi}$,$p_r$, and $r$ when $m_1=1.4\sunmass$ and when $m_1 =
    m_1 + \Delta m_1$. In each case, the other mass, $m_2$, of the
    system was fixed at $14.001\sunmass$. The {\tt SEOBNRv2\_opt} line, in
    each case, extends to about 5 orders of magnitude lower than the
    SEOBNRv2 line, before roundoff error becomes significant. The
    solid blue black diagonal line is the line of equal change in
    $m_1$ and the initial value.}
\end{figure}

Such sensitivities to initial conditions
are illustrated in Fig.~\ref{fig:Snstvtyeps}, where we plot the change
in these Hamiltonian input parameters as a logarithmic function of
perturbation added to one of the binary masses $m_1$. Notice that when
we replace the finite-difference derivatives of the SEOBNRv2
Hamiltonian with their exact expressions, computed using C-code generated
by Mathematica, the sensitivity to small perturbations in initial
conditions is improved by several orders of magnitude. We anticipate
these improvements in accuracy will have a positive impact on
using SEOBNR for PE algorithms that depend on taking small steps
through parameter space, such as Fisher-matrix-based estimates of uncertainties in the large signal limit. 

\section{Conclusions and Future Work}
\label{Conclusions}

In preparation for the 8-dimensional, precessing SEOBNRv3 (v3) model's
final incorporation into {\tt LALSuite}, we have worked to increase
the performance of the 4-dimensional SEOBNRv2 (v2) by $\sim$300x (Eq
\ref{eq:SpeedupFactor} and Table~\ref{tab:heat}), so that PE can now be
performed directly with SEOBNRv2 within weeks to months using
standard, serial-processing MCMC techniques, which could be reduced to
days or weeks with minimal parallelization. Our optimization
strategies have been shown to dramatically reduce the SEOBNRv2
run-time and over-sensitivity to initial conditions, while maintaining
amplitude-weighted phase agreement with the original code to within 0.00790 rad
(Table~\ref{tab:heat}), which is entirely dominated by roundoff
errors. The work discussed in this paper is hoped to not only assist
MCMC parameter estimation but be useful to others in the community who
cannot use pre-optimized SEOBNR codes due to their slow speeds (e.g.,
the generation of stochastic template banks \cite{Capano2016}). 

Some of our optimizations have already been applied to the SEOBNRv3
code within {\tt LALSuite}, leading to speed-ups of around 15x over
its original version. We anticipate that overall speed-ups of
$\sim$300x of v3 are likely once the remainder of our v2
optimizations have been incorporated, as the codes are identical in
structure and inefficiencies. 

As we work to optimize SEOBNRv3, the resulting v3 code can
be viewed as a stopgap for v3-based PE while efficient new 8-D ROM
strategies are invented, which may be capable of far faster waveform
generation than even a $\sim$300x-optimized SEOBNRv3 code. 

Excitingly, we do not believe even our optimized SEOBNRv2 code is
particularly efficient and are confident that significant, $\sim$100x 
optimizations are still possible (leading to overall speed-up factors
of $\sim$10,000x, over the original {\tt LALSuite} SEOBNRv2 and v3
codes). Though the release of SEOBNRv3 marks the end of our efforts
toward optimizing SEOBNRv2, we are eager to apply new optimization
ideas to v3, once all of the v2 optimizations have been
incorporated. With our planned optimizations of v3, a timely
interpretation of an observed gravitational wave using precessing
SEOBNR models directly may be within reach. 

\ack

We gratefully acknowledge A.~Buonanno, R.~Haas, P.~Kumar, and
A.~Taracchini for valuable discussions as this work was prepared. This
work was supported by NASA Grant 13-ATP13-0077 and NSF EPSCoR Grant
1458952.

\section*{References}

\sloppy

\bibliographystyle{unsrt}
\bibliography{references}{}

\end{document}